\documentclass[12pt,a4paper,final]{iopart}

\usepackage{iopams}  
\usepackage[breaklinks=true,colorlinks=true,linkcolor=blue,urlcolor=blue,citecolor=blue]{hyperref}
\usepackage{cite}
\usepackage{graphicx}
\usepackage{caption}
\usepackage{epsfig}
\usepackage{epstopdf}
\usepackage{textcomp}

\begin{document}

\title[Animation with Algodoo: a simple tool for teaching and learning physics]{Animation with Algodoo: a simple tool for teaching and learning physics}

\author{Samir L. da Silva$^{a}$$^{, 1}$; Rodrigo L. da Silva$^{b}$; Judismar T. Guaitolini Junior$^{a}$;
 Elias Gon\c{c}alves$^{a}$; Emilson R. Viana$^{c}$ and Jo\~{a}o B. L. Wyatt$^{a}$ }

\address{$^{a}$Instituto Federal do Esp\'{i}rito Santo, campus Vit\'{o}ria, 29040-780 Vit\'{o}ria, Brazil.\\
$^{b}$Instituto Federal Fluminense, campus Bom Jesus do Itabapoana, 28360-000 Bom Jesus do Itabapoana, Brazil.\\
$^{c}$Universidade Tecnol\'{o}gica Federal do Paran\'{a}, campus Curitiba, DAFIS, 80230-901. Curitiba, Brazil.\\
$^{1}$samir.lacerda@ifes.edu.br}

\begin{abstract}
In the present work we have made use of the animation freeware Algodoo, as an easy handling tool to teach and learn physics.  The animation is based on the oblique motion model and we have described the movement qualitatively, showing changes in the trajectory of an object as we modify the control parameters, such as speed and launch angle. Exploring the software graphical tools, the consistency between the results obtained by the animation and the literature is maintained, for example the maximum height and the rise time of the disc. This tool can be applied to students from different education levels, and also to undergraduate and high school students. 

\end{abstract}

\vspace{2pc}

\section{Introduction}
\label{sec1}

In the last decades, due to technological advancement, the numbers of tools that aim to facilitate teachers and students in the educational process have grown. Nowadays, an educator has the opportunity to demonstrate, in class, the evolution of physics equations and systems, by varying the parameters in real time, through commercial softwares as Wolfram Mathematica ~\cite{r01,r02}, Matlab ~\cite{r03} and Labview ~\cite{r04,r05,r06}. In other cases, simulations and/or animations ~\cite{r07,r08,r09,r10,r11} can be created, based on different computer languages, data acquisition boards to make automatization of experimental data ~\cite{r12,r13,r14} and software like Modellus ~\cite{r15,r16,r17,r18,r19,r20}, which is a free application that allows us to use mathematics to create or explore models interactively. Some authors used games ~\cite{r21} and console video games ~\cite{r22} to offer a new alternative for teaching physics.

The actual globalization process, Internet, technology, among others, are making the access to information a real resource for the learning process. Also, the velocity that one can obtain the information is somewhat incredible. Knowledge can be shared in blogs, social networks, video-classes, and even in virtual classrooms (as the Moodle platform ~\cite{r23,r24}). In this context, companies and teaching/research centers have adapted their products to meet this demand, allowing free access of licence sale.

Various are the tools used to support the teaching/learning process, but they still face resistance from a many users, since they require training and specific knowledge of computational programming skills. To avoid this situation, some companies have employed in their software accessibility protocols, to make the manipulation easier to any user. We can highlight the free simulation software Algodoo 2D (by Algoryx Simulation AB ~\cite{r25}). Algodoo presents a cartoon based environment, in which one can create scenarios using touch screen/mouse cursor, interacting with objects through clicks, drags, tilts and shakes. The software was developed to draw objects, which can be subjected to physical properties, and them simulated in a real environment such as gravity, air resistance, friction, elasticity, density, refractive index, forces, rotations, and so on. All those real world properties can be switched on and off, whenever necessary. It still provides inventions, games or simulations of physical systems, present in the set menus of the classroom. For a deeper analysis, the software offers graphics display of several variables, and also the visualization of vector variables in the trajectories of objects. When used on a computer with internet access, the Algodoo makes available to the users the opportunity to publish or access multiple projects in a free dynamical database. The software is available for Mac, Windows, and iPad platforms.

Another positive factor is that Algodoo encourages users of any age to explore their creativity, while the interaction with the software ambient is intuitive and relatively easy (when compared to advanced programm languages). The teacher has, with this tool, an interesting option to display the contents, and also a fun learning application for the students, that can be used at home, not only in the classroom.

In this paper, we present a classic physics problem in high school and undergraduate courses: the oblique motion. And it is addressed in the Algodoo environment, highlighting its easy manipulation. This study is organized in two stages: first we discuss a qualitative analysis of movement, emphasizing the trajectory of the object, while we control some parameters, in this case, the velocity magnitude and the launch angle. We also emphasize the modification of the velocity vector and its components along the motion. In the next step, we evaluated, through Algodoo's graphical tools, the observed effects in the first stage, by calculating the maximum height, the particle's range and the rise time. All simulations results will be compared to theory, by the equations of motion of the oblique motion. This tool can be applied to students from different education levels, and also to undergraduate and high school students. In our initial tests in Brazil classes, the students have shown a quick and better understanding of the physical content when they used the Algodoo's simulation. This simple theme has just been chosen to explore the software tools. The aim this paper is to use Algodoo for didactic purposes.

\vspace{0.3cm}
\section{The oblique motion model}
\label{sec2}

The oblique motion (OM) is a two-dimensional motion composed of two linear movements, one horizontal with constant speed and the other vertical with the constant acceleration of gravity, where here we are disregarding air movement and resistance, and also consider constant the gravitational force acting in the vertical direction near the planet's surface ~\cite{r26,r27,r28,r29}. The OM can be studied from kinematics and also Newton's force laws.

\begin{figure}[h!]
\begin{center}
\includegraphics[width=0.6\linewidth]{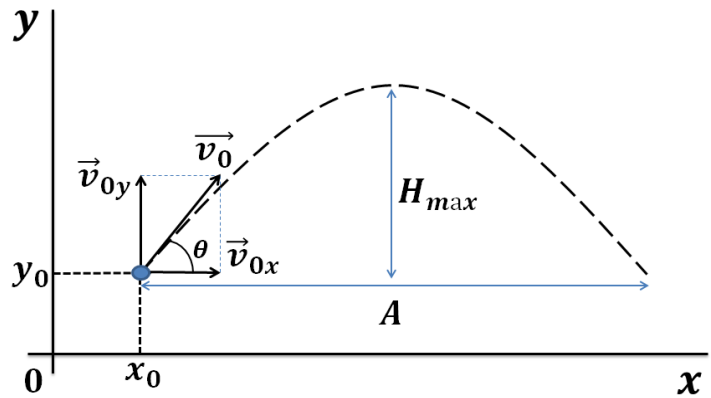}
\end{center}
\caption{Illustrative diagram of the oblique motion of a particle. The components of the initial velocity vector are ${v_{0x}=v_0\cos\theta}$ and ${v_{0y}=v_0\sin\theta}$. The initial position is given by the ordered pair ${(x_0,y_0)}$ relative to the origin O.}
\label{Figure1}
\end{figure}
\vspace*{4mm}
When a particle is launched under these conditions, with an initial velocity $\vec v_0$, not parallel to the acceleration of gravity, we observe a parabolic trajectory, where the position of the object, coordinates ${(x(t),y(t))}$, can be predicted by the parametric equations below:
\vspace*{3mm}

\begin{equation}
 x(t) = x_0+v_{0x}t
.\label{eq01}
\end{equation}

\begin{equation}
y(t) = y_0+v_{0y}t-\frac{1}{2}gt^2
.\label{eq02}
\end{equation}

\vspace*{2mm}
Here ${(x_0(t),y_0(t))}$ corresponds to the initial position of the object, and ${(v_{0x},v_{0y})}$ are the horizontal and vertical components of the initial velocity, respectively. The velocity makes an angle $\theta$ with the horizontal and g is the (constant) magnitude of the acceleration of gravity. This scheme is described in Figure 1.

The horizontal velocity ${v_{0x}}$ is constant throughout the trajectory, since there is no acceleration/force acting in this axis, and the vertical velocity component ${v_{0y}}$ varies in time, according to equation 3:
\vspace*{2mm}

\begin{equation}
v_y = v_{0y}-gt
.\label{eq03}
\end{equation}

\vspace*{2mm}
When the particle reaches the maximum height of the trajectory $H_{max}$, relative to the launch level, the value of the vertical velocity component becomes zero, i.e. $v_y = 0$ m/s. Exactly at this time, the vertical component of the velocity suffers a sign inversion. Manipulating equations (2) and (3) we obtain an expression for the maximum height:
\vspace*{2mm}

\begin{equation}
H_{max} = \frac{v_{0y}^2}{2g}
.\label{eq04}
\end{equation}

\vspace*{2mm}
The maximum horizontal distance traveled by the particle between the launch point and the drop point is called the range A. Since the time interval to achieve the range is twice the time to reach the maximum height (inasmuch as the trajectory is symmetric in the upward and downward stages), we can manipulate equations (1) and (3) to obtain an expression for the range A of the OM:
\vspace*{2mm}

\begin{equation}
A = \frac{v_{0}^2\sin{2\theta}}{g}
.\label{eq05}
\end{equation}

\vspace*{2mm}
The understanding of this model is the basis for the study of launches applications in sports, ballistic, rockets launching, for example. Given the initial conditions of the system, the main objective in these events is to provide the maximum height and the range attained by the object under analysis, or the reverse: through the maximum height and range obtains the initial conditions of launch.

\vspace{0.3cm}
\section{Teaching strategies}
\label{sec3}
\vspace{0.3cm}

\subsection{Qualitative analysis using Algodoo}
\label{subsec3.1}

In this work the 2D animation environment Algodoo, for Windows 7 platform, was used in a personal computer (PC), where studies on how the variables showed in the previous sections can control/modify the oblique motion. The main objects in the animations are drawings disks that can be differentiated by color. Each disk can assign different angles and launch velocities.

The discs drawn in Algodoo receive a speed v (vector) and an angle of launch $\theta$. In a frictionless environment, the disc describes a parabolic trajectory, as shown in Figure 2. Using the animation, the student can display the velocity vector and its components, verifying their magnitudes at any point of the trajectory. In Figure 2 we can see the effect of g (acceleration of gravity) on the orientation and magnitude of the velocity's vertical component along the trajectory. Furthermore, we note that the motion is symmetric with respect to the vertex of the parabolic trajectory, and also that the horizontal component of velocity remains constant throughout the route. These effects are predictable, by a detailed analysis of Equations (1) and (3).

The students should be encouraged by the teacher to change initial conditions, and them they will be able to observe what happens to the physical properties contained in the animation, in particular the values of position and speed on the trajectory.

\begin{figure}[h!]
\begin{center}
\includegraphics[width=0.6\linewidth]{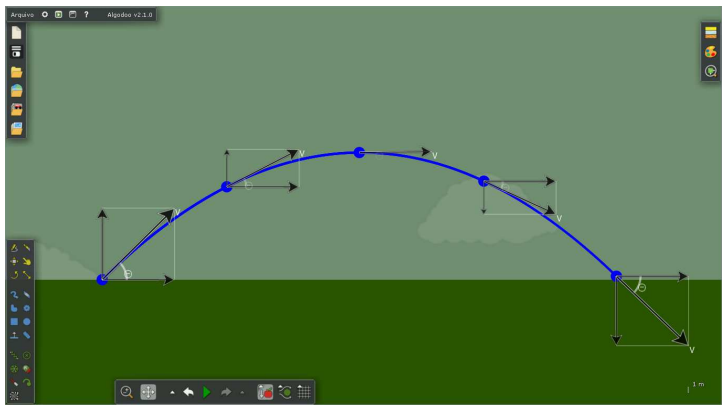}
\end{center}
\caption{Image of oblique motion a disc made in Algodoo. In the figure are shown the velocity of disc and its components at various positions along the trajectory.}
\label{Figure2}
\end{figure}

The software can also display the angle between the velocity $\vec{v}$ at any point of the trajectory and the acceleration of gravity. In Figure 2 the velocity magnitude is not present and its components do not overload the figure with too much information. The access to all data provides a more efficient study of the motion, by changing the control parameters. All this is achieved without writing a single line of programming, just by clicking the disk and selecting the values and orientations present in the box 'options' of the animation.

\begin{figure}[h!]
\begin{center}
\includegraphics[width=0.6\linewidth]{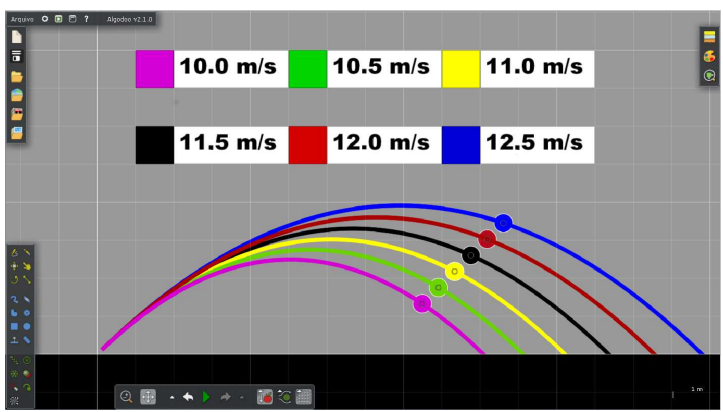}
\end{center}
\caption{Image of the oblique motions of disks with different initial velocities made in Algodoo. In the figure shown the magnitude of the speed a function of the color of the disc.}
\label{Figure3}
\end{figure}

The influence of the launch velocity magnitude in the trajectory of the disc can be seen in Figure 3, which illustrates a simulation where we have attributed different speeds, from 10 m/s to 12.5 m/s, for a set of six discs, each one indexed by different colors. We chose a launch angle of $45.0\,^{\circ}$ for all disks, drawing a trajectory corresponding to its color to start the animation. The results presented in Figure 3 are consistent with the prediction given by equations (4-5), in which, provided that the launch velocity increases, the range and the maximum height achieved by the disc also increases.

Now, assuming an initial speed with module equals 12.0 m/s, common to all disks, and assigning to each disk different launch angles, from $15\,^{\circ}$ to $75\,^{\circ}$, we have produced different ranges for the movement, as observed in Figure 4. In Equation 5, one can see that the maximum range $A_{max}$ occurs when the launch angle is $45\,^{\circ}$, and for larger values the disc reaches positions at ranges already visited, when launched with smaller angles. This fact can be portrayed in our animation in Algodoo, by the pair of trajectories with angles of $15\,^{\circ}$ and $75\,^{\circ}$ (pink and blue), and $30\,^{\circ}$ and $60\,^{\circ}$ (yellow and red), which present the same range for different launch angles. The maximum range is represented by the black trajectory that has a launch angle of $45\,^{\circ}$. A note can be added here: the ranges are equals for complementary angles (the sum equals $90\,^{\circ}$, for example $30\,^{\circ}$+$60\,^{\circ}$ = $90\,^{\circ}$); this can also be obtained from kinematics equations, and is an efficient exercise for students, for practicing the simulation and calculations.

\begin{figure}[h!]
\centering
\includegraphics[width=0.6\linewidth]{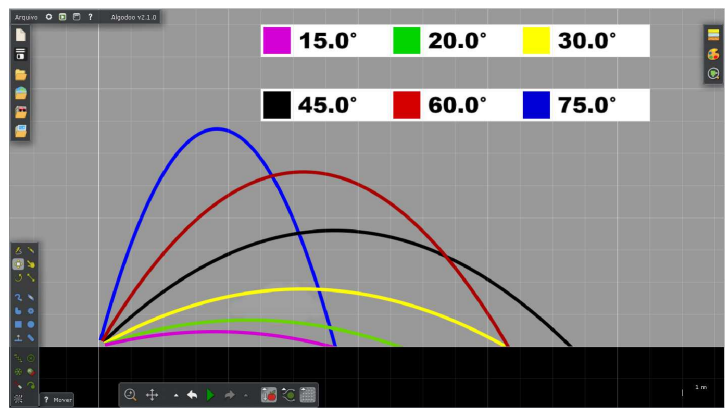}
\caption{Image of the oblique motions of disks with different launch angles made in Algodoo. In the figure shown the value of the angles a function of the color of the disc.}
\label{Figure4}
\end{figure}

\vspace{0.3cm}
\subsection{Quantitative analysis with Algodoo}
\label{subsec3.2}

In this section, we show how the teachers can introduce analysis and interpret graphs for position/speed/acceleration against time, by the use of Algodoo's graphs. All graphs can be seen simultaneously with the disc's trajectory, since both are constructed and drawed while the animation is running. At this moment, the teachers can promote debates concerning well-known student misconceptions about interpretations of this movement, and also contradiction between the position graph and the body trajectory ~\cite{r30,r31}.

The teacher has, in this moment, the tools to introduce the concept of composition of movement, exploring the two basic models discussed in the classroom, the uniform rectilinear motion (URM) and uniformly varied rectilinear motion (UVRM), and theirs corresponding graphical representation.

After analyzing qualitatively the projectiles motion in the previous section, we present in this section the animation of a disk launch with an angle of $45\,^{\circ}$ relative to the horizontal plane, with 15.0 m/s for the initial speed magnitude, in a region where is subjected to the action of gravity (acceleration g = 9.8\;m/$s^2$). The air resistance was disconsidered. The aim of this section is to use the Algodoo's graphical tools to determine the range, the maximum height and the time interval to achieve each of them. The graphical tools allow also that one can save the points to a file in .CSV format, or save the picture of the chart in .JPG.

\begin{figure}[h!]
\centering
\includegraphics[width=0.6\linewidth]{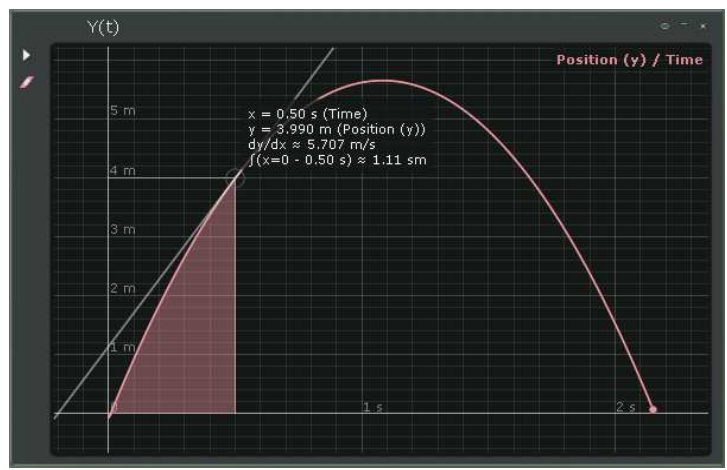}
\caption{The vertical position versus time in the oblique motion made in Algodoo. The figure represented the value of the ordered pair, the slope at the same point and the area bounded by the graph's curve and the horizontal axis.}
\label{Figure5}
\end{figure}

\begin{figure}[h!]
\centering
\includegraphics[width=0.6\linewidth]{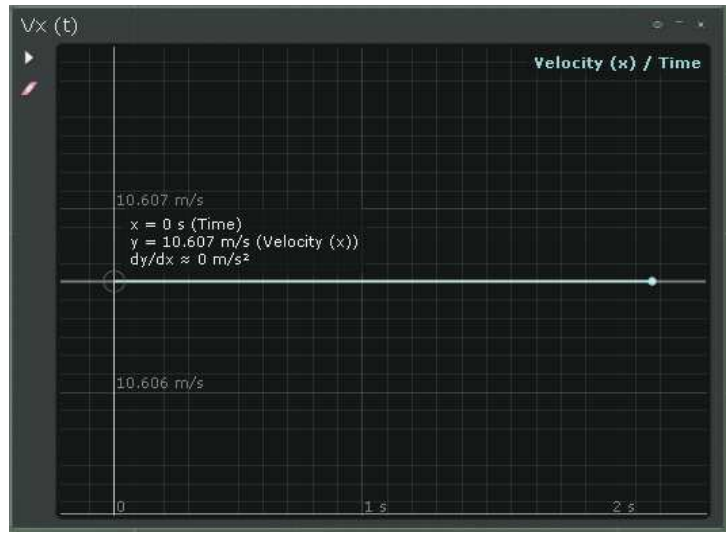}
\caption{Graph of horizontal component of velocity against time in the oblique motion in Algodoo. The figure presents a line parallel to the time axis indicating null acceleration.}
\label{Figure6}
\end{figure}

Another important point is the easy interaction with the graphical data in Algodoo environment. When the graph is done, we just need to touch the curve with the cursor (mouse or touching it), and the program draw a tangent line at that point. It also displays on the marked point the ordered coordinate pair, the value of the slope and the value of the area bounded by the curve and the horizontal axis of the graph (Figure 5). The slope is represented by an infinitesimal variation between vertical and horizontal variables, known as derivative $dy/dx$, and the area is given by the integral. These mathematical properties can be studied in a pray like Figure 5. The area drawn on the graphic of figure 5 has no physical meaning for the oblique motion, but this tool can be applied for other systems (by analyzing work done by a non-constant force, for example).

We now analyze the horizontal movement of the disc. In Figure 6 we show the horizontal component of the velocity $v_{0x}$ as a function of time. The time interval is measured for all graphs in this section, spanning from the moment of launch until the moment when the disk touches the ground.

The area, in Figure 6, between the curve and the time axis has now physical meaning and indicates the spatial variation in the x-axis, $\Delta x$. Thus, in Figure 6 we calculated the area and determine $A = 22.952$ m, which has an error of only 0.03\% relative to the value of 22.959 m predicted by the equation 5. 

In Figure 7 we show the graph of the horizontal position a function of time, $x(t)$. In this graph, the teacher can introduce derivative concepts, since there are known that any tangent line to the space-time graph indicates the instantaneous velocity at that point. Thus Figure 7 confirms that the speed is constant for any time instant, since the slope of the tangent line is constant (the curve is a straight line). By calculating the graph of the spatial variation $\Delta x$, one can also find the range of the movement: in the example of the Figures $A = 22.981$ m, with 0.096\% error. The error in the simulation measure of A using Figure 7 is three times larger than those calculated by Figure 6, since the disk size is relevant to define its exact position at each instant of time. To minimize this error, the size of the disk should be reduced.

\begin{figure}[h!]
\begin{center}
\includegraphics[width=0.6\linewidth]{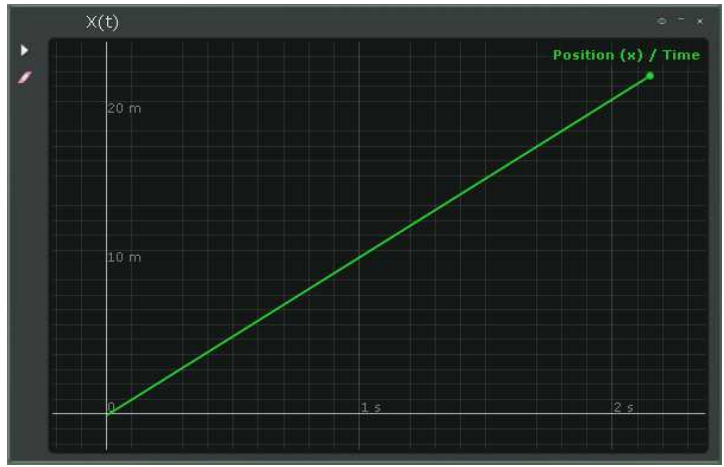}
\end{center}
\caption{Graph of horizontal component of position against time in the oblique motion in Algodoo. The figure shows a straight line with a slope of ${v_{0x}=10.607\;m/s}$.}
\label{Figure7}
\end{figure}

Teachers should encourage their students to compare the horizontal motion described in Figures 6 and 7 with the uniform rectilinear motion (URM) discussed in class.

The vertical component of the velocity $v_y$ as function of time is presented in Figure 8. The slope of the curve indicates the direction and magnitude of the gravity acceleration (g), and we obtain the constant value $g = - 9.80\;m/s^2$. Also by studying Figure 8, one can determine the instant of time that the disk reaches the maximum height, by touching/clicking the cursor on the curve at the instant when the velocity is zero. The value obtained for the initial velocity used in this example - $v_0 = 15.0\; m/s$ - was 1.08 seconds. Measuring the area under the curve from 0.00 s to 1.08 s, the value of the vertical displacement corresponding to the maximum height can be obtained: $H_{max} = 5.73\; m$. The value provided by Equation 4 is 5.74 m with an error of 0.17\%. 

The area under the curve in Figure 8 represents the vertical displacement of the disc, and in this case the total displacement is zero. This illustrates the symmetry of this kind of movement, in which the negative and positive portions of the displacement obtained by graphic are identical in module. Therefore, the object back to launch level at the end of the trajectory, for the present example. One can also study the projectile movement by analyzing when the object of study reaches different levels: below the launch level or above it.
\vspace*{2mm}

\begin{figure}[h!]
\begin{center}
\includegraphics[width=0.6\linewidth]{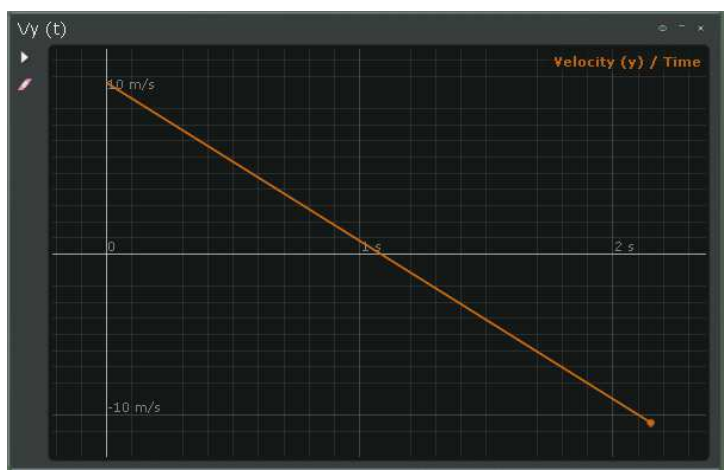}
\end{center}
\caption{Graph of vertical component of velocity as time function in the oblique motion in Algodoo. The figure shows a straight line with a slope of ${ g=-9.8\; m/s^2}$.}
\label{Figure8}
\end{figure}

Finally, Figure 9 shows the vertical position as a function of time $y(t)$. The graph is a parabola with concave down as predicted by Equation 2. At absolute maximum, i.e., the y-coordinate of the vertex, we find the value of the maximum height $H_{max} = 5.74\; m$. This result is identical (up to millimeters) to that obtained using Equation 4. Still exploring the graph given by the Algodoo animation, we observe that the slope of the tangent line to the curve indicates the direction and magnitude of the vertical velocity $v_y$, at each point. We obtain positive and decreasing values for the slopes when we observe the slopes from the launching point to the parabola's vertex. The positive value indicates that the disk is rising (positive y-coordinate) and, due to the action of gravity, it is decelerating, since the magnitude of the y-velocity is decreasing, until $v_y$ becomes zero at the vertex of the curve. After the absolute maximum of $y(t)$, we obtain negative and increasing values for the velocity (also by measuring the slopes of the curve at each point). In other words, this procedure confirms that after vertex the disc is falling, uniformly increasing the module of its velocity. The student can observe, from experience, the relation between acceleration and velocity; also he/she can analyze how the mathematical signs of the acceleration and velocity connect each other.

The students should compare the vertical motion described in Figures 8 and 9 with the uniformly varied rectilinear motion (UVRM) discussed in the classroom. Teachers can ask students to find the equation of the horizontal and vertical movement of the disc, through the initial conditions: speed and position. The parameters (maximum range and flight time) can be obtained easily from the graph using Algodoo. The teacher or the student may also analyze the data in a .CSV file, for example, on other software, to calculate the average speed and acceleration, establishing the conditions for which it is physically acceptable to consider the average values as instantaneous speed and acceleration of discs.

\begin{figure}[h]
\begin{center}
\includegraphics[width=0.6\linewidth]{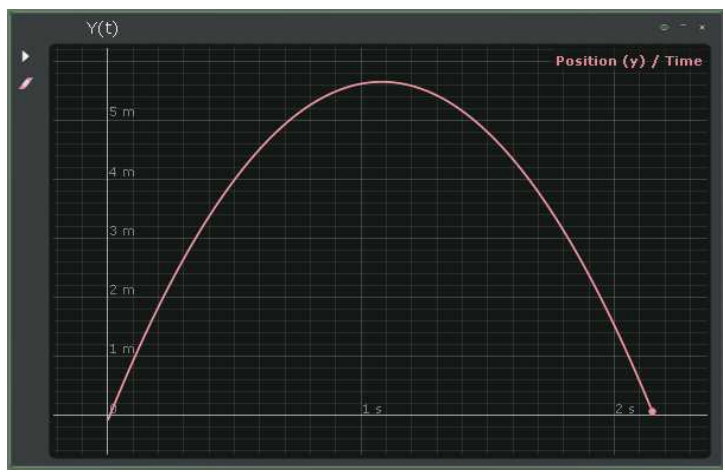}
\end{center}
\caption{Graph of vertical component of position as time function in the oblique motion in Algodoo. The figure shows a parabola with concave down.}
\label{Figure9}
\end{figure}
\vspace*{2mm}

This section presents only a simple although important example of kinematics, which can be exploited by the teacher in the classroom with computational animations using the freeware Algodoo, an easy handling tool. The animations using Algodoo can also be exploited in several other topics in physics education, and it is up to the teacher to establish scripts to work these animation tools with their students.

\vspace{0.3cm}
\section{Conclusions}
\label{sec4}

In this paper, we show a computational tool of great potential for teaching and learning physics, using the freeware Algodoo. Using an animation based on the oblique model, we describe the motion demonstrating qualitatively the changes in the trajectory due to modifications of the control parameters, such as initial velocity and launch angle. Exploring the Algodoo's graphical tools, we present coherence between the results of the maximum height and the rise time of the disc, obtained graphically or by the theory, equations (1-5). The errors obtained in the graphs can also be studied in class, since its sources are comprehensible. This tool provides a good understanding of the kinematics contents, where the teacher may promote discussions with their students during the animation. It's worth noting that, since technology is extremely accessible nowadays, useful freeware that can be used as academic resource, are gaining field in classrooms, and this paper contributes with an easy and interactive animation for physics teaching/learning.

We demonstrated the use of a simple animation to study the oblique motion that is independent of specific programming knowledge, since is easy to accomplish the tasks in this software. With this environment the educator and the student can explore the full potential of the studied subject, proposing modifications in the system such changings in the acceleration of gravity or in the level differences between the departure or arrival points. Within the modification of the departure/arrival points, the students may realize that not always ready equations are directly applicable, as those routinely used to compute fall time and range in parabolic motion, depending on the symmetry of the system and initial speeds. Occurring this computational intuitive perception, the software will help the teacher to show that the most important feature in the learning process is to interpret the equations and physical system, applied for each particular situation, instead of memorizing formulas.

Thus, the Algodoo is a helper tool that puts the scenario of teaching and learning in physics easier and enjoyable, when compared to others platforms and animation education ~\cite{r01,r02,r03,r04,r05,r06,r07,r08,r09,r10,r11,r12,r13,r14,r15,r16,r17,r18,r19,r20,r21,r22,r23}. The teaching of physics using this freeware can be applied to students of different ages and level of education. The Algodoo provides the study of several physical systems creatively and interactively, and this research line promotes a new topic for the future of education.

\vspace{0.3cm}
\section{Acknowledgments}
\label{sec4}

This work is supported in part by FAPERJ, Brazilian official agency.

\end{document}